 \documentclass[iop]{emulateapj}

\shortauthors{Straughn et al.}

\usepackage{subfigure}
\usepackage{natbib}
\newcommand{\etal}	{\mbox{et al.\,}}

\newcommand{\Mo}	{\mbox{M$_{\odot}$}}

\begin{document}

\title{A Multiwavelength Study of Tadpole Galaxies in the Hubble Ultra Deep Field}

\author{
Amber N. Straughn\altaffilmark{1}, Elysse N. Voyer\altaffilmark{2}, Rafael T. Eufrasio\altaffilmark{1,3}, Duilia de Mello\altaffilmark{4}, Sara Petty\altaffilmark{5}, Susan Kassin\altaffilmark{6}, Jonathan P. Gardner\altaffilmark{1}, Swara Ravindranath\altaffilmark{6}, Emmaris Soto\altaffilmark{4}
}

\altaffiltext{1}{Astrophysics Science Division, Goddard Space Flight Center, Code 665, Greenbelt, MD 20771, USA}
\altaffiltext{2}{Randstad at Google, 1129 San Antonio Road, Palo Alto, CA, USA}
\altaffiltext{3}{Department of Physics and Astronomy, Johns Hopkins University, Homewood Campus, Baltimore, MD 21218, USA}
\altaffiltext{4}{Department of Physics, The Catholic University of America, Washington, DC 20064, USA}
\altaffiltext{5}{Department of Physics, Virginia Tech, Blacksburg, VA 24061, USA}
\altaffiltext{6}{Space Telescope Science Institute, 3700 San Martin Drive, Baltimore, MD 21218, USA}

\begin{abstract}

Multiwavelength data are essential in order to provide a complete picture of galaxy evolution and to inform studies of galaxies' morphological properties across cosmic time.  Here we present results of a multiwavelength investigation of the morphologies of ``tadpole'' galaxies at intermediate redshift (0.314 $<$ $z$ $<$  3.175)  in the Hubble Ultra Deep Field.  These galaxies were previously selected from deep Hubble Space Telescope (HST) F775W data based on their distinct asymmetric knot-plus-tail morphologies (Straughn \etal 2006).  Here we use deep Wide Field Camera 3 near--infrared imaging in addition to the HST optical data in order to study the rest--frame UV/optical morphologies of these galaxies across the redshift range 0.3$<$z$<$3.2.  This study reveals that the majority of these galaxies \emph{do} retain their general asymmetric morphology in the rest--frame optical over this redshift range, if not the distinct ``tadpole'' shape.  The average stellar mass of tadpole galaxies is lower than field galaxies, with the effect being slightly greater at higher redshift within the errors.  Estimated from SED fits, the average age of tadpole galaxies is younger than field galaxies in the lower redshift bin, and the average metallicity is lower (whereas the specific star formation rate for tadpoles is roughly the same as field galaxies across the redshift range probed here).  These average effects combined support the conclusion that this subset of galaxies is in an active phase of assembly, either late--stage merging or cold gas accretion causing localized clumpy star--formation.  

\end{abstract}


%%%%%%%%%%%%%%%%%%%%%%%%%  INTRO: SEC. 1 %%%%%%%%%%%%%%%%%%%%%%%%%
\section{Introduction}
Galaxies in the local universe are well--studied and adequately described by the Hubble sequence, with their morphologies largely correlating with their physical properties.  As we look to higher redshifts, the picture becomes more complex, but similar correlations persist and galaxy morphology continues to provide important insights into the overall process of galaxy evolution.  The advent of large, deep, multiwavelength datasets allows us to advance our knowledge about galaxies at higher redshifts, and provides new ways of studying different galaxy types across a large range of wavelength.  In particular, galaxy surveys that probe the redshift range $\sim$1$<$z$<$3 provide insight into a very important phase of galaxy evolution, encompassing the peak of cosmic star formation when galaxies were actively forming stars and assembling their mass (Lilly \etal 1996; Madau \etal 1998; Hopkins \& Beacom 2006).  This redshift range also encompasses a ``transition'' redshift (around z$\sim$1.7--2), when galaxies as a whole were changing in terms of morphology, star formation processes, gas accretion, and galaxy growth in general (Lehnert \etal 2015, Cameron \etal 2010, Driver \etal 2013).

Multiwavelength data have become increasingly important for studying galaxies at these intermediate redshifts, providing a more complete picture of galaxy evolution.  This applies to morphological studies as well as deriving physical properties from the galaxies' spectral energy distributions (SEDs).  UV light directly probes recent star formation from young, massive stars whereas optical light allows investigation of underlying older stellar populations while still probing massive ionizing stars (e.g., Calzetti 2007, Bond \etal 2014).  In order to study these properties in galaxies at higher redshift, deep near--infrared imaging is required to study rest--frame UV and optical features.  Hubble Space Telescope (HST) galaxy surveys (the deep fields, COSMOS, GOODS, GEMS, AEGIS, and most recently, CANDELS) have enabled extensive multiwavelength studies of galaxies across cosmic time.

Morphological and structural studies of galaxies have traditionally been undertaken in two broad regimes: visual classifications and quantitative classifications---both parametric (e.g., surface brightness profiles) and non--parametric (CAS parameters, Conselice 2003; Gini/$\mathrm{M_{20}}$, Abraham \etal 2003, Lotz \etal 2004).  Automated quantitative classifications have the obvious advantage of being able to handle large datasets quickly.  Although these methods generally work reasonably well at low redshift, quantitative classifications become much more difficult at redshifts higher than $z\sim$1.  For example, simulations of nearby galaxies to higher redshifts (i.e., ``artificially redshifted galaxies'') provide a glimpse into the difficulties of automatically detecting faint features at redshifts above $z\sim$1.  Automated methods such as Gini, $\mathrm{M_{20}}$, and concentration are able to detect asymmetries and distinguish between disks and multiple nuclei (e.g., Conselice \etal 2004, Lotz \etal 2006, Law \etal. 2007, Petty \etal 2009, de Mello \etal 2006). However, detecting structural details such tidal tails, low surface brightness disks, and clumps prove more difficult.  The difficulty in detecting and classifying such structures in high redshift galaxies is largely due to the surface brightness dimming of extended features such as tidal tails.  A decrease in spatial resolution also contributes to fewer individual clumps such as massive starburst regions being identified.  This effect is particularly noticeable when using automated algorithms such as Gini and $\mathrm{M_{20}}$, where these effects start at redshifts as low as $z=$0.5 (Petty \etal 2014).

In general, when studying galaxy morphologies, high signal--to--noise data is critical for detecting and studying galaxies with intrinsically faint structures, such as tidal tails which are often present in interacting galaxies.  This was made evident with the earliest HST deep fields, in which an abundance of distant galaxies was seen with complex structures (e.g., Driver \etal 1995; Steidel \etal 1996; Ellis \etal 1998); subsequent deeper imaging confirmed the ubiquitous nature of galaxies of complex morphology at higher redshift (Conselice \etal 2004; Elmegreen \etal 2004, Lotz \etal 2006; van den Bergh \etal 2002).  The Hubble Ultra Deep Field (HUDF) revealed galaxies of various morphological types (e.g., Elmegreen \etal 2005) including a subset classified as ``tadpole'' galaxies defined by a clump of bright star formation plus an extended, diffuse ``tail'' (Elmegreen \etal 2005; Straughn \etal 2006, Elmegreen \etal 2010).  These tadpole galaxies are a subset of a larger collection of irregular and clumpy galaxies defined by star--forming clumpy structure that is common at high redshift (Guo \etal 2015).

Tadpole galaxies as a subset have proven interesting to study in part due to the relative ease of selecting their morphologies; i.e., a bright star--forming ``head'' offset to a diffuse ``tail.''  There are several possibilities of the origin of the morphology, including edge--on disks with a large star--forming region at one end, asymmetrical ring galaxies viewed edge--on (Elmegreen \& Elmegreen 2010), or some type of interaction or assembly process (Straughn \etal 2006, Windhorst \etal 2006, Rhoads \etal 2005).  In the deepest imaging available, tadpoles at moderate redshift represent somewhere between 6--10\% of galaxies depending on size cuts (Straughn \etal 2006, Elmegreen \etal 2005).  Whereas this morphology is fairly common at high redshift, it is much more rare in the local universe (Elmegreen \etal 2012; Sanchez Almeida \etal 2013).  Local tadpoles provide an interesting laboratory to study these types of galaxies in detail.  Elmegreen \etal (2012) found that heads of local tadpoles are comparable to other local star--forming regions and the local tadpole tails resemble bulge--less disks.  Sanchez Almeida \etal (2013) measured H$\alpha$ velocity curves and abundances of seven local tadpoles, and found complex rotation dynamics and metallicity gradients more similar to high--redshift disk galaxies than local disks.  They interpret young ages and very low metallicities in the tadpoles' heads as resulting from metal--poor gas inflows.

%%%%%%%%%%%%%%%%%%%%%%%%%  DATA & SAMPLE SELECTION: SEC. 2 %%%%%%%%%%%%%%%%%%%%%%%%%
\section{Data \& Sample Selection}
The sample of tadpole galaxies presented here is drawn from the catalog in Straughn \etal (2006).  This parent catalog relied on an automated detection procedure to identify tadpole galaxies from the Hubble Ultra Deep Field (HUDF) Advanced Camera for Surveys (ACS) \emph{i'}--band (F775W) image.  The motivation behind the automated detection procedure was to attempt to reduce the bias from purely visual morphological selection techniques.

This detection procedure was based on varying the SourcExtractor (Bertin \& Arnouts 1996) deblending parameter and selecting on the ellipticity of sources in order to detect compact, circular sources (tadpole ``heads'') in nearby spatial proximity to elongated diffuse sources (tadpole ``tails'').  In particular, the SourcExtractor deblending parameter determines whether peaks in flux are considered part of one source, or are deblended into multiple sources.  With this parameter set to a high value, spatially nearby flux maxima are separated into different sources; conversely, with the deblending parameter set to a low value, spatially nearby peaks in flux are considered part of the same source.  Using the highly--deblended SourcExtractor output catalog with a circular ellipticity selection produces compact tadpole ``heads'', and using the low deblended catalog with a high ellipticity cut produces the tadpole ``tails'' catalog.  Subsequently the heads were matched with the tails, and failures in the procedure were discarded (14 obvious misdetections were omitted from the initial selection).  In this way, objects are selected based on physical peaks in flux, rather than a purely visual method which is subject to biases based on how the data is displayed.  Straughn \etal (2006) provides a complete description of the details of this detection procedure and the resulting initial sample. 

In the current paper, our broad aim is to investigate the morphologies and physical properties of the tadpole galaxies using the latest deep near--infrared imaging data, combined with the deep ACS HUDF data used in the previous study.  We use the publicly--available Wide Field Camera 3 (WFC3) data from the Hubble Ultra Deep Field.  These combined full--depth mosaics are from the HUDF09 project (HST Program ID 11563, PI: G. Illingworth; Bouwens \etal 2011), the HUDF12 project (HST Prorgram ID 12498, PI: R. Ellis; Ellis \etal 2013, Koekemoer \etal 2013), the CANDELS project (HST Program ID 12060,12061,12062; PI: S. Faber, H. Ferguson; Koekemoer \etal 2011, Grogin \etal 2011), and HST Program ID 12099 (PI: A. Riess).

The ACS HUDF \emph{i'}--band (F775W) data (Beckwith \etal 2006) used in the original study has a depth of 144 orbits (Straughn \etal 2006 sample was selected to F775W 28.0 mag).  The near--infrared data used here are combined mosaics of the programs described above, with combined depths as follows: F105W: 100 orbits; F140W: 30 orbits; F160W: 84 orbits (see Fig. 1).

The area of the WFC3/IR data is smaller than the ACS (\emph{i'}--band) HUDF area on which the Straughn \etal (2006) study was based, due to the smaller field of view of WFC3/IR compared to the ACS/WFC (by a factor of 2.4).  This resulted in a reduction of the initial sample from 165 galaxies to 81.  Since the \emph{i'}--band automatically--selected galaxies from the previous study were selected from deeper and higher resolution data (in which the distinct tadpole morphology was easily resolved in most cases), that study reached a much fainter magnitude limit.  Many of the faintest of these galaxies were not significantly detected at all in the WFC3/NIR data (and were not in the catalogs of the above programs), or had surface brightnesses that were too low and/or the galaxies were too small to be significantly detected and visually classified in the NIR (resolution effects are discussed in Section 5).  After removing these sources due to lack of usable data, the final sample contains 26 tadpole galaxies as defined in Straughn \etal (2006) to a limiting \emph{i'}--band magnitude of 25.8 mag (corresponding H--band limit of 25.5; see Table 1).  A visual examination of the deep NIR data produced no new tadpole candidates (i.e., there were no NIR--identified tadpole galaxies that were not already identified in the F775W data).

%%%%%%%%%%%  PHYSICAL PROPERTIES OF TADPOLE GALAXIES: SEC. 3 %%%%%%%%%%%%%%%%%%%%%%%%%
\section{Physical Properties of Tadpole Galaxies}

%%%%%%%%%%%%%%   3.1 PHYSICAL PROPERTIES  %%%%%%%%%%%%%

A gallery of tadpole images is shown in Figure 1; individual images are 3" x 3'' and each row displays the F775W, F160W, and combined four--color images to highlight morphologies across wavelength of these sources.  The single--band images show some of the differences in morphology across wavelength and also clearly shows the effect of resolution on morphology determination.  To further investigate these effects, we performed a resolution matching exercise by convolving the F775W data to the F160W PSF; the results of this exercise are presented in Section 5, and the resulting convolved image cutout stamps are shown as the fourth stamp in each row in Figure 1.

%%%%%%%%%  FIG. 1: STAMPS %%%%%%%%%%

\begin{figure*}
%\hspace{-6.6cm}
%%\includegraphics[scale=0.8]{ev_tadpole_gallery1_V2_photz.eps}
\hspace{1.9cm}
\includegraphics[width=7cm]{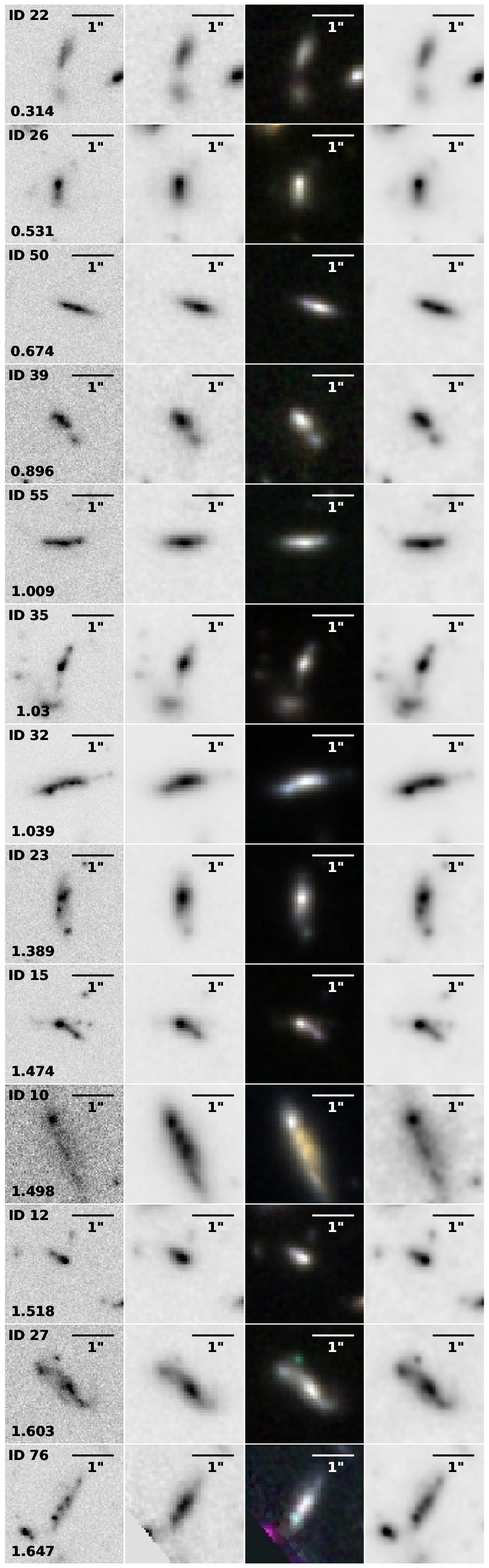}
%\hspace{-9cm}
%%\includegraphics[scale=0.8]{ev_tadpole_gallery2_V2_photz.eps}
\includegraphics[width=7cm]{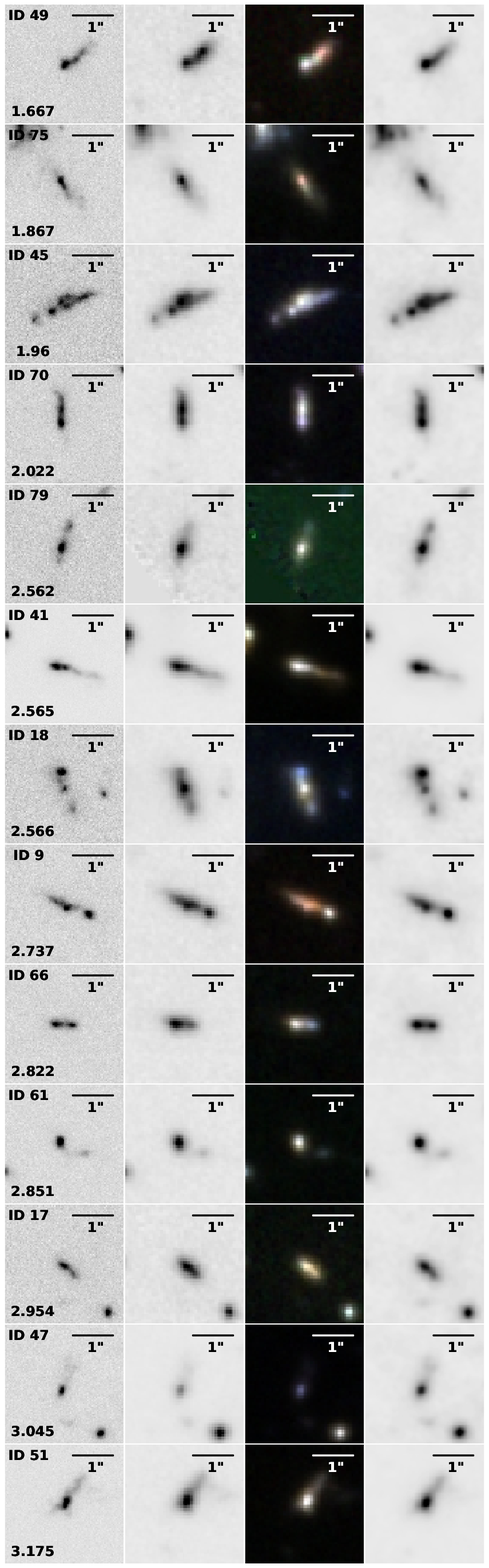}
%\hspace{-10cm}
%\caption{}
\caption{Fig. 1---Gallery of tadpole galaxies arranged in order of increasing redshift; redshifts are given in lower--left corner.  Panels shown for each galaxy are F775W, F160W, 4--color composite (F775W, F105W, F140W, and F160W), and convolved images (F775W data convolved with F160W PSF).} 
\end{figure*}

The CANDELS project provides the most recent photometry (Guo \etal 2013) and photometric redshifts (T. Dahlen \etal in prep., Santini \etal 2015; Dahlen \etal 2013 describes the technique used to determine photometric redshifts) for the tadpole galaxies in our sample.  Several of the tadpole galaxies have ground--based spectroscopic and/or grism--spectroscopic redshifts; however, all of these were of intermediate to low quality.  Therefore, the most reliable redshift estimates for all of the galaxies in this sample were determined to be the photometric redshifts.  The redshift range of the sample is 0.314 $<$ $z$ $<$  3.175 (Figure 2) with an average redshift of $z_{avg}=$1.75 and median redshift of $z_{med}=$1.57.  

%%%%%%%%%  FIG. 2: PHOTZ DISTRIBUTION %%%%%%%%%%
\begin{figure}
\includegraphics[scale=0.5]{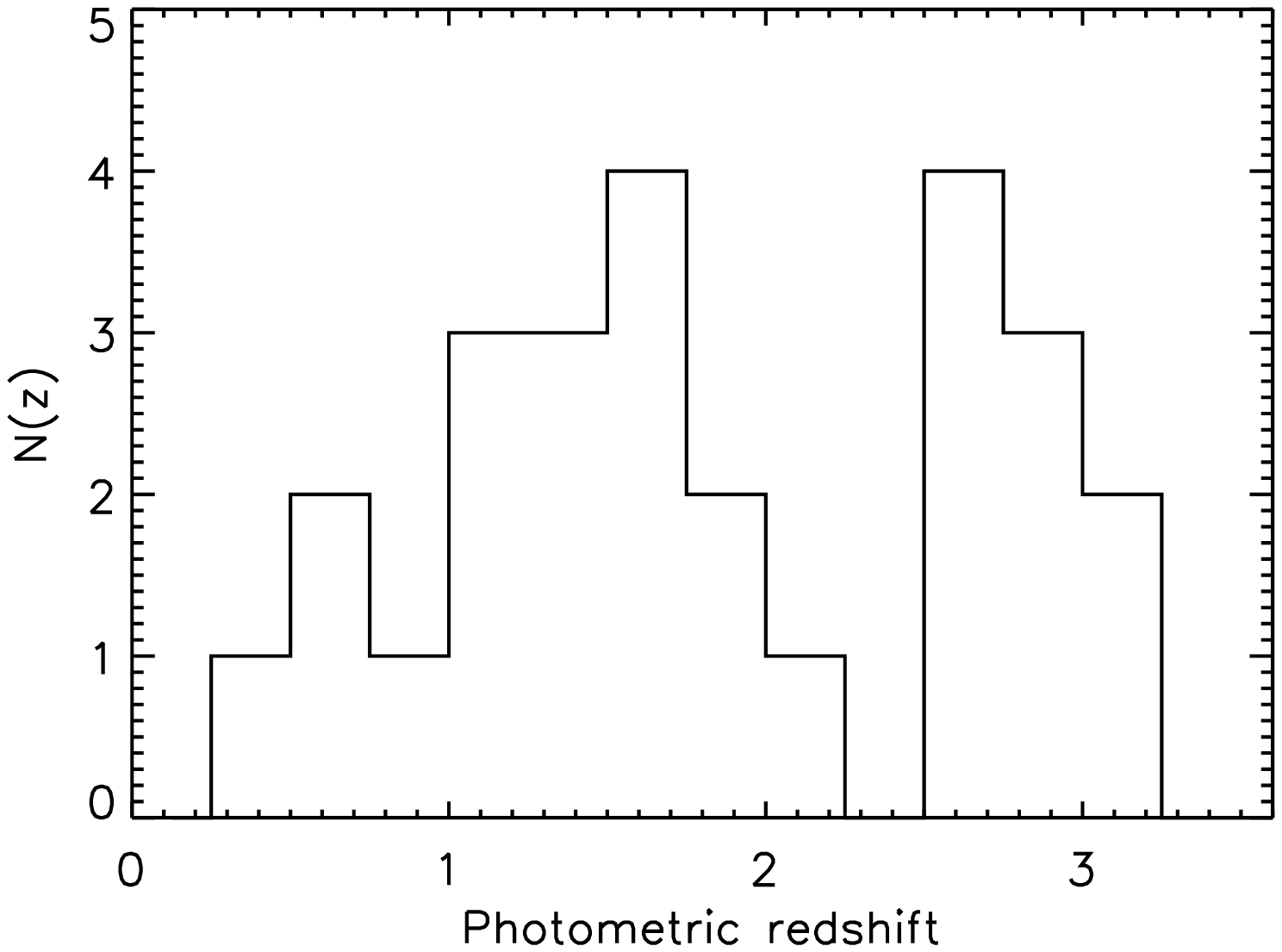}
\caption{Photometric redshift distribution for the tadpole galaxy sample; this sample spans a redshift range of 0.3$<$z$<$3.2.}
\end{figure}

Using CANDELS data, Santini \etal (2014) have completed an extensive and robust study of determining galaxy stellar masses and other physical parameters from a variety of SED--fitting methods.  In particular, they adopt a median approach from nine different SED--fitting methods in GOODS-South in order to greatly reduce systematics and arrive at improved stellar mass estimates.  Here we use this reference median stellar mass (see Table 1), as well as other physical parameters derived from the Santini \etal 2014 study (age, SFR, metallicity).  We caution here that while the stellar masses used here are as reliable as possible given the data (and methods used to arrive at them), the other physical parameter estimates derived from SED fits still suffer from degeneracies, particularly age and metallicity.  However, dividing the tadpole sample into two redshift bins (z$\le$2 and z$\ge$2) allows us to obtain a more broad interpretation of the physical processes in these galaxies.  This is a natural dividing line, as there is evidence that z$\sim$2 is an important transition redshift in terms of galaxy morphology, star formation, and galaxy growth in general (e.g., Cameron \etal 2010, Lehnert \etal 2015).  When investigating the physical parameters of the tadpole galaxies in comparison to the field galaxies in the same area of the HUDF, we restrict the comparison to galaxies to the same brightness limit as the tadpoles (H$\le$25.5).

The tadpole galaxies' average masses in each bin are shown in Figure 3, with HUDF field galaxies to the same magnitude limit shown for comparison.  The tadpoles' stellar masses are lower in both bins to within the errors (mass being the only parameter for which errors are provided in Santini \etal 2014), with the effect being slightly greater at higher redshift.  The tadpole galaxies' average ages in the two redshift bins are shown in Figure 4.  The average ages of both samples increase with time (with decreasing redshift) as expected.  For the low--redshift bin, the average age of tadpoles is lower than field galaxies, and in the high--redshift bin, it is roughly the same as the average age of the field galaxies.  This is as expected; at earlier cosmic times, the cosmic star formation rate density was higher and galaxies' average stellar populations on average were younger.  By redshift $\sim$1, the cosmic star formation rate density had very much decreased \emph{on average}; however, isolated galaxy populations (such as the tadpole galaxies) were still experiencing ongoing star formation, yielding younger stellar populations.  As expected, the specific star formation rate (sSFR) of the sample increases as a function of redshift, along with the field sample (Figure 5); however, the tadpoles do not show significantly enhanced sSFRs compared to the field sample.  The metallicity of the tadpoles compared to the field are shown in Figure 5.  As expected, both populations show increasing metallicity with time.  For both redshift bins, the tadpole sample has lower metallicity than the field sample.

%%%%%%%%%  FIG. 3: MASS DISTRIBUTION BINNED %%%%%%%%%%
\begin{figure}
\includegraphics[scale=0.5]{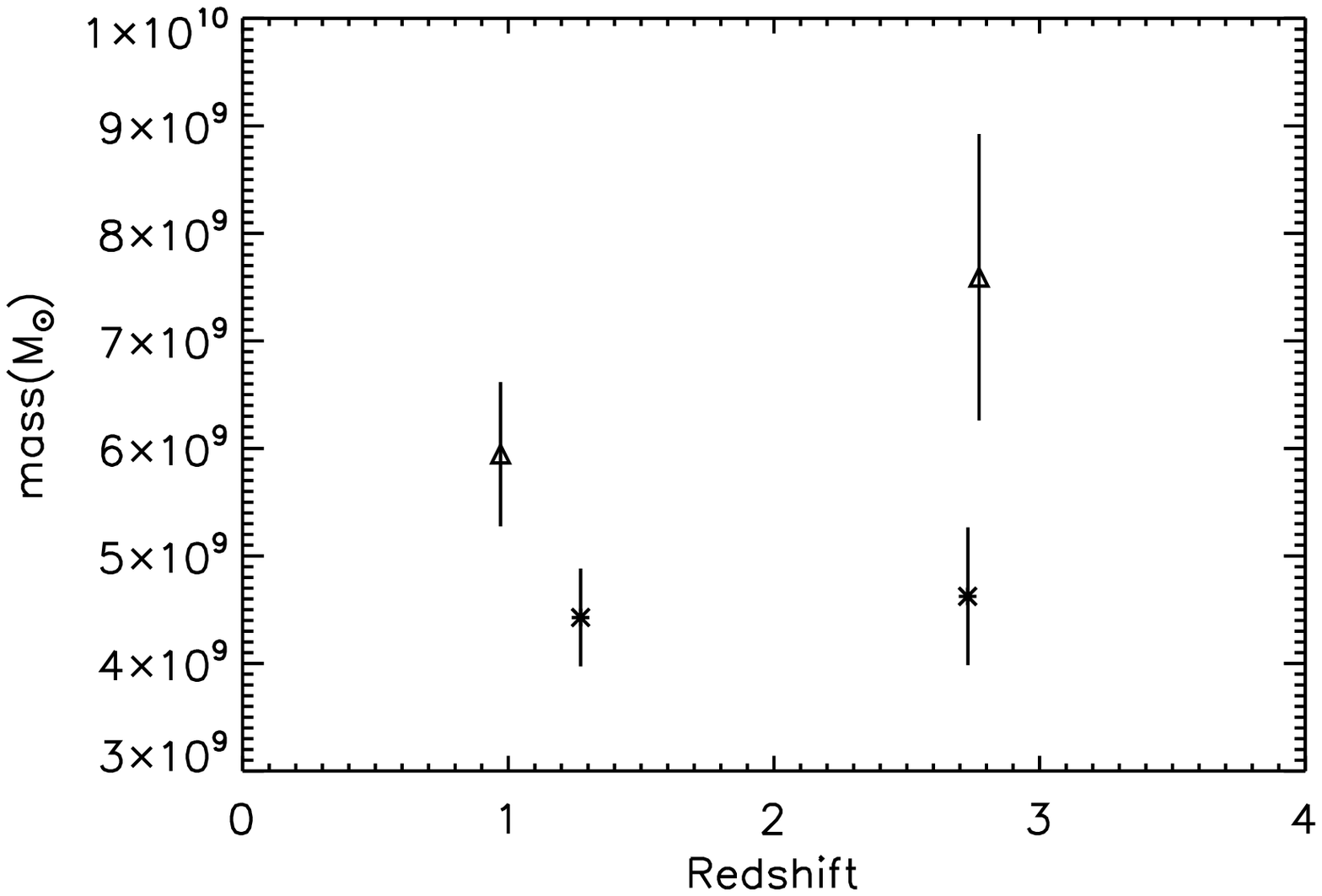}
\caption{Average mass of tadpole galaxies (stars) and HUDF field galaxies (triangles) to the same magnitude limit (H$\le$25.5) in high-- and low--redshift bins.  At both z$<$2 and z$>$2, tadpoles have lower stellar masses within the errors, with the effect being slightly greater at higher redshift.}
\end{figure}

%%%%%%%%%  FIG. 4: AGE AS FCN OF REDSHIFT %%%%%%%%%%
\begin{figure}
\includegraphics[scale=0.5]{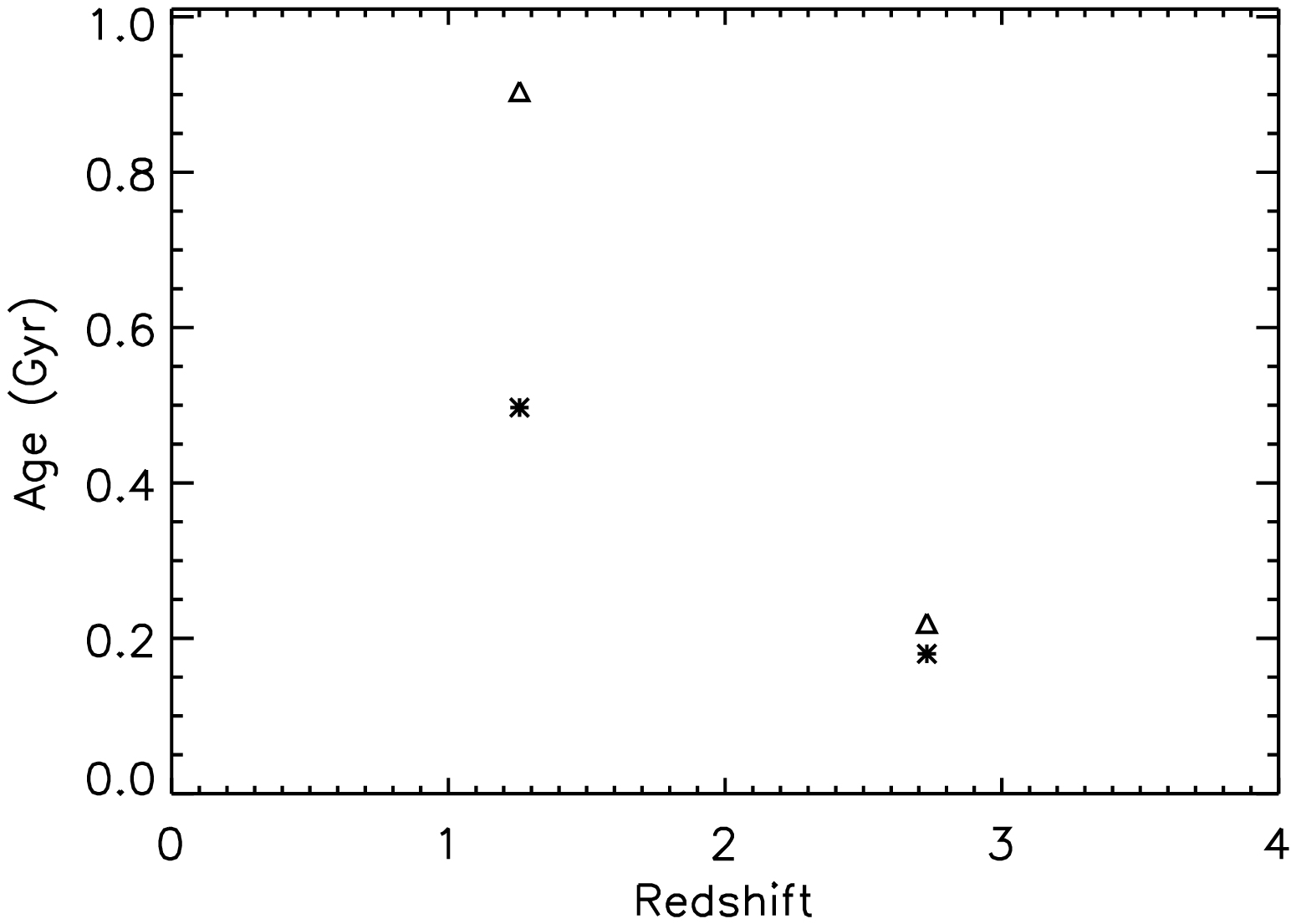}
\caption{Average age of tadpole galaxies (stars) in high-- and low--redshift bins, compared to the average age of HUDF field galaxies (triangles) to the same magnitude limit.  The average age of tadpoles is lower by $\sim$400Myr in the low--redshift bin, whereas the average ages of both samples are roughly the same in the high--redshift bin.}
\end{figure}

Sanchez Almeida \etal (2013) show that local low--mass, young tadpole galaxies are comparatively low metallicity, with some tadpole ``heads'' having extremely low metallicity ($<$0.1 Solar), and very young ages ($<$5 Myr).  Combined with complex dynamical properties, they attribute the local tadpole galaxies' properties to accretion of metal--poor gas from the ISM.  The emerging picture from the present sample of tadpole galaxies--a population that compared to the field sample is on average less massive, younger, and lower metallicity--supports this interpretation.

%%%%%%%%%  FIG. 5: SSFR %%%%%%%%%%
\begin{figure}
\includegraphics[scale=0.5]{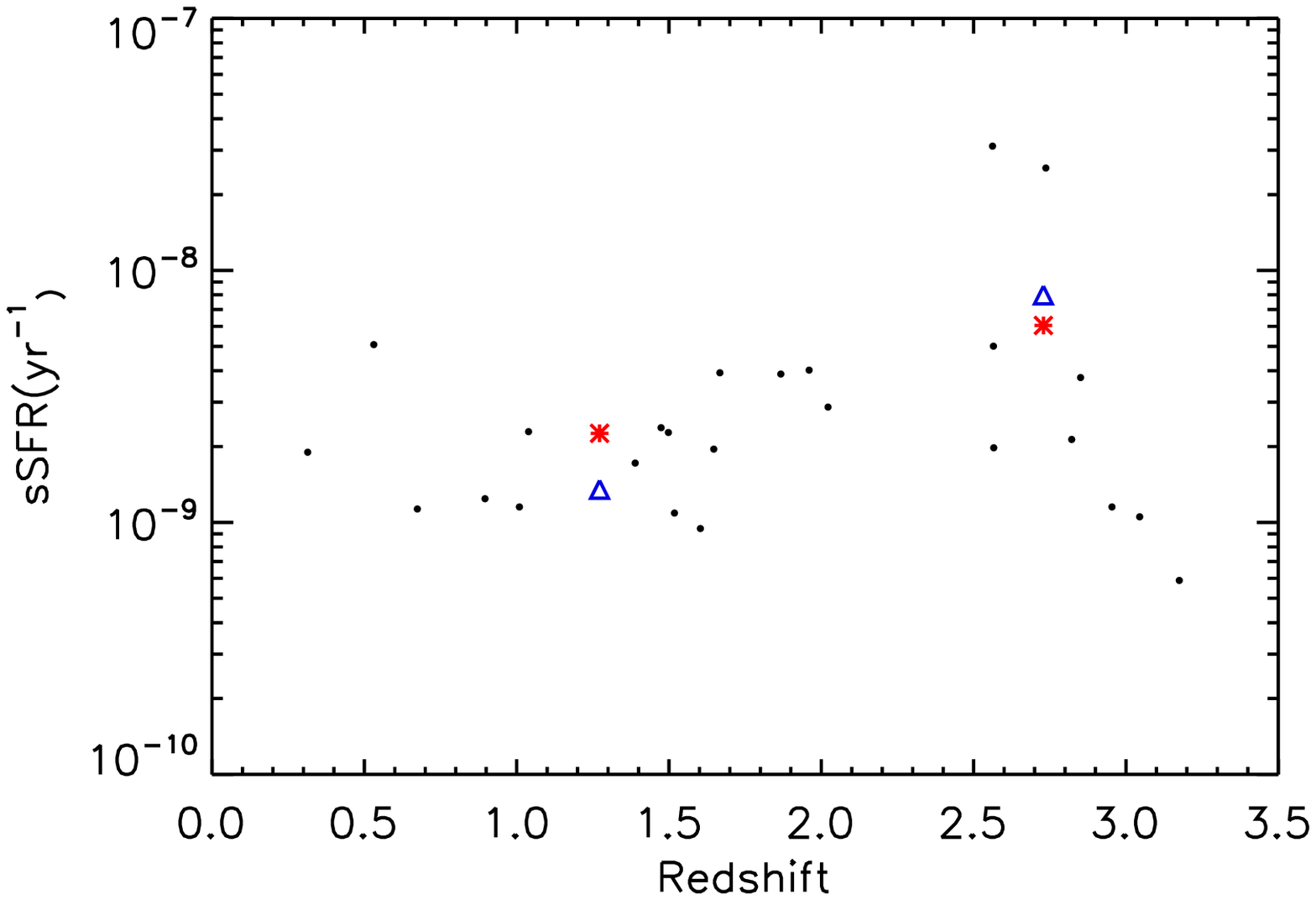}
\caption{Specific star formation rate of tadpoles as a function of redshift.  The entire sample is shown as filled black circles, with the low-- and high--redshift bin averages shown as red stars.  For comparison, the HUDF field sample averages in each bin are show to the same brightness limit.}
\end{figure}

%%%%%%%%%  FIG. 5: METALLICITY AS FCN OF REDSHIFT %%%%%%%%%%
\begin{figure}
\includegraphics[scale=0.5]{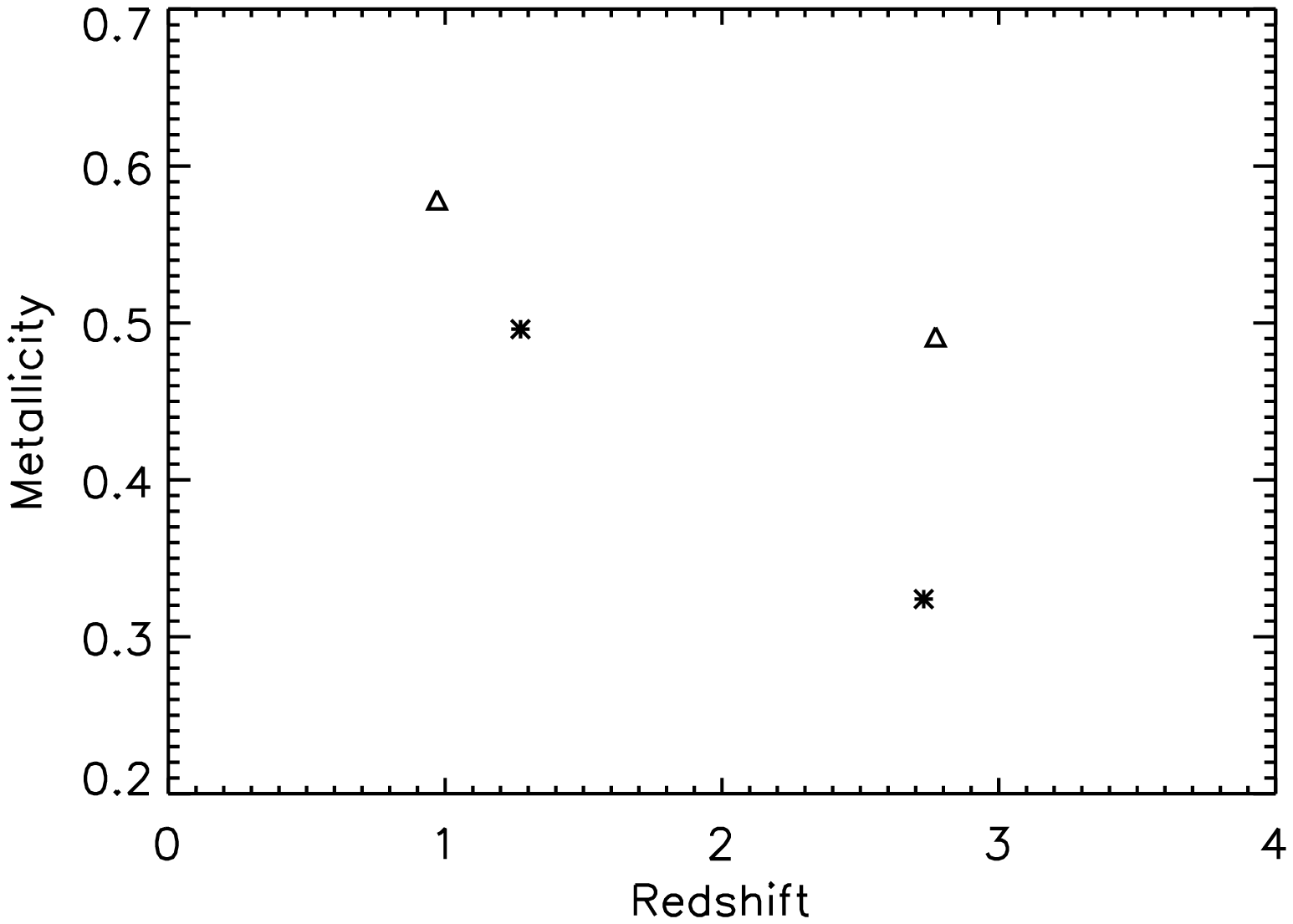}
\caption{Average metallicity of tadpole galaxies (stars) in high-- and low--redshift bins, compared to the average metallicity of HUDF field galaxies (triangles) to the same magnitude limit.  The average metallicity of tadpoles is lower in both redshift bins.}
\end{figure}

%%%%%%%%%%%%%%   4. MORPHOLOGIES  %%%%%%%%%%%%%

\section{Multiwavelength Visual Morphologies of Tadpole Galaxies}

Galaxy morphology studies are both important and difficult; quantitative/automated methods often fail at low S/N, high redshift, and/or low surface brightness, leaving visual morphology classifications as the standard for many studies especially at high redshift.  Previous study of this tadpole galaxy sample (Straughn \etal 2006) relied solely on the HST/ACS optical data available at that time; here we aim to investigate any change in visual morphology of the tadpole galaxies across wavelength.  Deep HST/WFC3 imaging allows us to look at the tadpoles' rest-frame optical and UV morphology at higher redshift.  For this relatively small sample, a qualitative look (Figure 1) shows that many of the F775W selected sample does appear to retain at least some indication of asymmetric morphology.  However, more structured, robust visual classifications of galaxy morphology are useful.  To that end, we use the CANDELS team morphology results (Kartaltepe \etal 2014) for the tadpole sample.  Kartaltepe \etal (2014) provides visual classifications of 2534 galaxies in the deep area of GOODS--South to H$\le$24.5 mag.  For each galaxy classified, the catalog lists the fraction of classifiers that selected certain criteria: main morphological classification in the H--band (disk, spheroid, and/or irregular), interaction classes, flags for specific morphologies (including tadpoles and asymmetry, as well as a flag for k--correction), and an estimation of clumpiness and/or patchiness (see Kartaltepe \etal 2014 for a full description).    

The CANDELS morphology study magnitude limit is almost a full magnitude brighter than that used in this study (H$\le$25.5), therefore not all of the tadpole galaxies are included in the CANDELS classifications.  A total of 15 of the tadpole galaxies are classified by the CANDELS team; all of these galaxies had at least five classifiers.  The main morphological classification of the CANDELS sample was performed on the H--band images.  The summary of the CANDELS visual morphology study as it relates to the tadpole sample in this paper is: 1.) Tadpole galaxies are generally classified as ``disks'' in the H--band at all redshifts, with some z$\ge$2 tadpoles classified as ``irregulars''; 2.) tadpole galaxies are generally flagged as ``asymmetric'', though not generally flagged as ``tadpoles'', ``chains'', or ``mergers'', and 3.) interactions have high levels of disagreement among visual classifiers.

A closer look at the CANDELS visual morphology conclusions as they relate to the tadpole sample reveals some interesting results.  First, the most basic effect here is that the CANDELS classifications were performed in the H--band (whereas the tadpoles were selected in the \emph{i'}--band).  The lower resolution of the H--band data causes some of the more compact star--forming clumps detected in the rest--frame optical or UV to not be well resolved in the NIR data.  The majority of the tadpole sample was classified as ``disk'' in the H--band; however, two tadpoles (both at z$>$2) had their main H--band CANDELS classification selected equally among classifiers as irregular and disk (IDs 18 and 41).  In the CANDELS classification scheme, among the possible flags was one for k--corrections based on the main morphological classification.  In the entire sample from Kartaltepe \etal (2015), only a small fraction ($\sim$1\%) had more than half of the classifiers check this flag; i.e., only a small fraction are classified as having significantly different morphologies in the V-- and H--bands.  None of the tadpole sample had this flag checked by more than half of the classifiers.  

Second, the majority of the tadpole galaxies had the ``asymmetric'' flag checked by more than half of the classifiers (and all of the galaxies had the ``asymmetric'' flag checked by at least one classifier).  However, the specific ``tadpole'' flag was \emph{not} generally checked (only three sources had at least half the classifiers check this flag---IDs 15, 41, 51).  This is also possibly the result of the galaxies being classified in the H--band, where some compact morphological properties are not resolved in the NIR.  This result does point to the conclusion that while morphological features might not be distinct enough to warrant a specific morphological classification (such as ``tadpole'' or '``chain''), the asymmetric characteristics of these types of galaxies does persist through wavelength and to higher redshift.  Therefore, asymmetry could be an important classification factor in selecting future samples of these types of galaxies at higher redshift, even when the detailed morphological properties cannot be resolved.  

Third, the CANDELS morphology classification exercise included flags for interactions.  One interpretation of tadpole galaxies is that they are interacting, although there are other interpretations for galaxies of this morphological type as well (propagating star--formation, Papaderos \etal 1998; gas instabilities in forming disks, Elmegreen \etal 2005, 2012).  From the CANDELS classifications, two tadpoles had the majority of classifiers select the ``any interaction'' flag (IDs 41, 76).  However, whether or not a galaxy is undergoing an interaction is one of the flags with the highest level of disagreement, in this subsample and also in the general CANDELS sample (Kartaltepe \etal 2015).  For example, for this sample, 73\% of the tadpoles had at least one classifier check the ``any interaction'' flag; however, 93\% of the sample also had at least one classifier check ``none'' for the interaction class.  This is inherent in the subjective nature of visual classifications (especially for galaxies without a ``simple'' morphological structure) even when an extensive training set is used in the beginning of the exercise. 

\section{Resolution Matching}

The question then arises: if this sample of tadpole galaxies was initially selected based on the distinct knot--plus--tail morphology in the \emph{i}--band, and the galaxies are generally \emph{not} classified specifically as ÒtadpolesÓ in the H-band, what is causing the difference between passbands? There are two broad possible answers: genuine differences in morphological
structures in different wavelengths (i.e., hidden bulges appearing in the H--band or star--forming rest--UV bright clumps disappearing in the longer wavelengths), or changes due to lower resolution in the near-IR data (such as distinct clumps becoming more blended at lower resolution), or some combination of these two effects.

In order to investigate the relative importance of these two effects, we have convolved the F775W data to the F160W PSF with the following basic question in mind: are the tadpole galaxies in the F775W image convolved to the F160W PSF more likely to retain the irregular and/or tadpole shape than the F160W images of the same galaxies?  If the answer is yes, then that suggests that genuine changes in morphology across wavelength are occurring (and the degraded resolution does not have significant impact on distinct morphological features); if the answer is no, then resolution effects are comparatively more important (the morphological features remain, but are smoothed out).  The native FWHM resolution of the F775W and F160W are $\sim 0.08$\arcsec\ and $\sim 0.15$\arcsec, respectively. We have modeled the PSFs in grids of 0.01\arcsec\ of both maps by measuring point sources across the field and circularly symmetrizing them (as described in detail in Aniano et al. 2011).  We have produced a convolution kernel to convolve the F775W data to the spatial resolution of the F160W map, using the model PSFs of both maps and following the procedure described in Gordon \etal (2008). The resulting F775W map was constructed to exactly match the spatial resolution of the F160W, allowing a direct morphological comparison of the two maps.

In Figure 1, the last image stamp in each row shows the F775W image convolved to the F160W PSF.  In order to investigate the question posed above, six of the authors classified the convolved images of the tadpole subset that was classified by the CANDELS team as part of the HUDF, using the same CANDELS visual classification scheme.  Random field galaxies from the convolved image were included in the set in attempt to reduce bias.  We note that a similar number of people classified HUDF galaxies in the general CANDELS classification (5 or 6; see Section 4).  The main results of this exercise---in comparison to the main results from the CANDELS H--band classifications of these galaxies---are as follows.  First, whereas the majority of the tadpole galaxies were classified as ``disk'' in the H--band, the majority of tadpoles are classified as ``irregular'' in the convolved image.  Second, the majority of the convolved tadpoles had the asymmetric flag checked by more than half of the classifiers, which is the same result as the H--band image.  However, most of the convolved galaxies also had the ``tadpole'' flag checked, which was not the case for the H--band (note, however, our discussion of bias below).  For interactions, three convolved tadpole galaxies had the majority of classifiers select the ``any interaction'' flag, whereas two tadpoles had that flag selected in the H--band; however, these were not the same galaxies in the two different cases.  Similarly to the H--band, most of the tadpoles had at least one classifier check the ``any interaction'' flag, while all but three galaxies had at least one classifier check ``none'' for the interaction class.  Half of the classifiers checked the ``merger'' flag for the convolved tadpole sample.

The fact that the convolved galaxies were generally classified as ``irregular'' rather than ``disk'' point to genuine morphological changes across wavelength (clumps remain in the convolved i--band) rather than degraded resolution; e.g., the clumps weren't smoothed out enough to cause the galaxies to be classified as disks.  Similarly, the fact that the ``tadpole'' flag was checked suggest that the clumps remain as distinct morphological features despite the reduced resolution.  This exercise also confirms what was found for this sample in the H--band classifications: there is high disagreement concerning classification of interactions in these galaxies.  The general results of this exercise appear to point to genuine changes in morphology across wavelength being significant for this sample, at least as much as resolution effects alone.  We emphasize here, that while we attempted to reduce bias in this exercise by adding convolved field galaxies to the sample, that there is still likely some bias present in this exercise since the authors knew that the focus of this work is tadpole galaxies specifically: e.g., the prevalence of the ``tadpole'' flag being checked is certainly subject to this bias.  However, we conclude that the overall results of this exercise support the idea that structural differences across wavelength are important to morphological classifications.

%%%%%%%%%%%%%%%%%%%%%%%%%  Summary: SEC. 6 %%%%%%%%%%%%%%%%%%%%%%%%%
\section{Summary}

The tadpole galaxies presented here at 0.314 $<$ $z$ $<$  3.175 represent a subset of clumpy galaxies spanning the peak of cosmic star formation and provide insight into the importance that multiwavelength data play in discerning galaxy properties out to intermediate redshift.  The morphologically--disturbed tadpole galaxies are generally younger, less massive, and have lower metallicities than field galaxies to the same brightness limit, qualitatively supporting the idea that they are in the process of assembly, either through a merger event or gas accretion, similar to results of local tadpole galaxies (Sanchez Almeida \etal 2013).  Selected from the deep, high--resolution F775W HUDF data based on their distinct knot--plus--tail morphology, these galaxies generally do retain their asymmetric morphology in the deep WFC3/NIR data based on visual morphological classifications from the CANDELS team, if not the distinct ``tadpole--like'' features.  Changes in morphological features across wavelength appear to play a more significant role than resolution effects in this sample of galaxies.

Detailed study of distant galaxy samples with low surface--brightness features such as the tadpole galaxies requires imaging depth on par with the Hubble Ultra Deep Fields.  In order to increase sample sizes and push studies of these types of galaxies to higher redshift higher resolution data in the infrared is needed, which will be provided with the upcoming James Webb Space Telescope.

{\it Facilities:} \facility{HST (ACS, WFC3)}

%%%%%%%%%%%%%%%%%%%%%%%%%  BIBLIOGRAPHY %%%%%%%%%%%%%%%%%%%%%%%%%

%%%%%%%%%%%%  TABLE 1  %%%%%%%%%%%%%%%%%%
\begin{deluxetable}{cccccccccc}
%\rotate
\tabletypesize{\scriptsize}
\tablecaption{Properties of Tadpole Galaxies \label{table1}}
\tablewidth{0pt}
\tablehead{
\colhead{ID} & \colhead{RA} & \colhead{Dec} & \colhead{AB (F775W)} & \colhead{AB (F160W)} & \colhead{Redshift} & \colhead{mass} & \colhead{mass error} & \colhead{age} & \colhead{SFR} \\
\colhead{} & \colhead{(deg)} & \colhead{(deg)} & \colhead{(mag)} & \colhead{(mag)} & \colhead{(photz)} & \colhead{\Mo} & \colhead{\Mo} & \colhead{Gyr} & \colhead{\Mo/yr}
}
\startdata
  9 & 53.1428871 & -27.7798996 & 25.33 & 24.37 & 2.7370 & 3.09e+09 & 9.49e+08 & 7.40 & 7.878e+01 \\
10 & 53.1434822 & -27.7831783 & 25.38 & 22.33 & 1.4980 & 5.15e+10 & 1.05e+10 & 8.65 & 1.171e+02 \\
12 & 53.1449394 & -27.7900829 & 25.55 & 24.72 & 1.5180 & 1.01e+09 & 2.12e+08 & 9.00 & 1.100e+00 \\
15 & 53.1468964 & -27.7817497 & 25.00 & 24.17 & 1.4740 & 9.68e+08 & 1.80e+08 & 8.65 & 2.300e+00 \\
17 & 53.1486092 & -27.7799168 & 25.77 & 25.26 & 2.9540 & 2.04e+09 & 5.10e+08 & 8.35 & 2.350e+00 \\
18 & 53.1499405 & -27.7904186 & 25.28 & 23.75 & 2.5660 & 1.40e+10 & 3.28e+09 & 8.90 & 2.769e+01 \\
22 & 53.1512451 & -27.7895470 & 25.65 & 25.48 & 0.3140 & 1.58e+07 & 4.70e+06 & 9.00 & 3.000e-02 \\
23 & 53.1525421 & -27.8003845 & 24.90 & 23.63 & 1.3890 & 2.50e+09 & 4.56e+08 & 8.90 & 4.300e+00 \\
26 & 53.1535797 & -27.7677803 & 24.57 & 23.98 & 0.5310 & 1.93e+08 & 6.50e+07 & 8.00 & 9.800e-01 \\
27 & 53.1538239 & -27.7763233 & 24.85 & 23.87 & 1.6030 & 2.20e+09 & 4.04e+08 & 8.75 & 2.080e+00 \\
32 & 53.1558609 & -27.7948952 & 23.48 & 22.76 & 1.0390 & 2.57e+09 & 3.72e+08 & 8.75 & 5.890e+00 \\
35 & 53.1566200 & -27.7942944 & 24.96 & -99.0 & 1.0300 & -9.000 & -9.000 & -9.000 & -9.000 \\
39 & 53.1581116 & -27.7925358 & 25.79 & 24.78 & 0.8960 & 3.06e+08 & 7.01e+07 & 8.85 & 3.800e-01 \\
41 & 53.1600990 & -27.7763138 & 23.93 & 23.45 & 2.5650 & 5.07e+09 & 1.17e+09 & 8.30 & 2.538e+01 \\
45 & 53.1638451 & -27.7653122 & 25.08 & 24.27 & 1.9600 & 1.43e+09 & 4.07e+08 & 8.40 & 5.750e+00 \\
47 & 53.1641808 & -27.7728844 & 25.71 & 24.78 & 3.0450 & 3.61e+09 & 1.77e+09 & 8.40 & 3.800e+00 \\
49 & 53.1646843 & -27.7943611 & 25.18 & 25.05 & 1.6670 & 4.84e+08 & 1.02e+08 & 8.40 & 1.900e+00 \\
50 & 53.1647301 & -27.7680111 & 25.43 & 25.28 & 0.6740 & 1.15e+08 & 4.28e+07 & 9.00 & 1.300e-01 \\
51 & 53.1649742 & -27.7651615 & 24.39 & 23.92 & 3.1750 & 9.67e+09 & 2.61e+09 & 8.50 & 5.690e+00 \\
55 & 53.1674805 & -27.7674618 & 25.36 & 24.73 & 1.0090 & 4.34e+08 & 7.90e+07 & 9.00 & 5.000e-01 \\
61 & 53.1723824 & -27.7939129 & 25.43 & 24.96 & 2.8510 & 1.36e+09 & 5.05e+08 & 8.35 & 5.110e+00 \\
66 & 53.1739998 & -27.7910118 & 25.52 & 24.78 & 2.8220 & 2.31e+09 & 4.79e+08 & 8.60 & 4.930e+00 \\
\enddata
\tablenotetext{*}{NOTE: full photometry can be found in Guo \etal (2014).  Redshifts here are the best measured, from T. Dahlen \etal in prep; in this case, all photometric redshifts.  Masses, ages, and SFRs all from Santini \etal (2014).}
%\tablenotetext{\dag}{CDF-S X-ray sources identified as AGN by Szokoly \etal 2004.}
%\tablenotetext{\ddag}{New spectroscopic redshift (no previous photometric or spectroscopic redshift measurements).}
\end{deluxetable}

%% Tables should be submitted one per page, so put a \clearpage before
%% each one.

%% Two options are available to the author for producing tables:  the
%% deluxetable environment provided by the AASTeX package or the LaTeX
%% table environment.  Use of deluxetable is preferred.
%%

%% Three table samples follow, two marked up in the deluxetable environment,
%% one marked up as a LaTeX table.

%% In this first example, note that the \tabletypesize{}
%% command has been used to reduce the font size of the table.
%% We also use the \rotate command to rotate the table to
%% landscape orientation since it is very wide even at the
%% reduced font size.
%%
%% Note also that the \label command needs to be placed
%% inside the \tablecaption.

%% This table also includes a table comment indicating that the full
%% version will be available in machine-readable format in the electronic
%% edition.

\end{document}